\documentclass[11pt]{article}
\usepackage{latexsym}
\usepackage{amsmath}
\usepackage{amscd}
\usepackage{graphics}
\usepackage{psfig}
\graphicspath{{epsfigures/}} 
\def\be#1\ee{\begin{equation}#1\end{equation}} 
\def\bea#1\eea{\begin{eqnarray}#1\end{eqnarray}} 
\newcommand{\la}{\lambda}
\newcommand{\mip}{\!-\!p}
\newcommand{\nmion}{n\! -\!\!\, 1}
\newcommand{\nmila}{n\! -\!\!\,\lambda}
\newcommand{\lamion}{\lambda\! -\!\!\,1}
\newcommand{\De}{D_\epsilon}
\newcommand{\cDe}{{\mathcal{D}}_\epsilon}

\newcommand{\cDela}{{\mathcal D}_\epsilon^\lambda}

\newcommand{\cDenmila}{{\mathcal D}_\epsilon^{n\! 
-\!\!\,\lambda}}

\newcommand{\Qd}{Q_\delta}
\newcommand{\Deprime}{D_{\epsilon'}}
\newcommand{\Deprimela}{D_{\epsilon'}^\lambda}
\newcommand{\Deprimenmila}{D_{\epsilon'}^{n\! -\!\!\,\lambda}}
\newcommand{\cDeprime}{{\mathcal D}_{\epsilon'}}
\newcommand{\cDeprimela}{{\mathcal D}_{\epsilon'}^\lambda}
\newcommand{\cDeprimenmila}{{\mathcal D}_{\epsilon'}^{n\! 
-\!\!\,\lambda}}
\newcommand{\cN}{{\mathcal N}}
\newcommand{\cNla}{{\mathcal N}^\lambda}
\newcommand{\cNnmila}{{\mathcal N}^{n\! -\!\!\,\lambda}}
\newcommand{\Nla}{N^\lambda}
\newcommand{\Nnmila}{N^{n\! -\!\!\,\lambda}}
\newcommand{\Aeta}{A_\eta}
\newcommand{\Beta}{B_\eta}
\newcommand{\pd}{\partial}
\newcommand{\ovl}{\overline}
\newcommand{\mathR}{{\rm I\! R}}

\newcommand{\cM}{\mathcal{M}} 
\newcommand{\cE}{\mathcal{E}}
\newcommand{\cU}{\mathcal{U}} 
\newcommand{\cV}{\mathcal{V}}

\newcommand{\cD}{\mathcal{D}}
\newcommand{\rar}{\rightarrow} 
\newcommand{\Rar}{\Rightarrow}
 
\newcommand{\cint}{\int}

\newcommand{\cP}{{\mathcal{P}}}
\newcommand{\cF}{{\mathcal{F}}}

\newcounter{def}
\newcommand{\definition}[1]{
\refstepcounter{def}\begin{list}{}{
\setlength{\leftmargin}{6mm}\setlength{\parindent}{0mm}}\item
{\bf Definition \thedef {~~}}{#1}\end{list}}

\newtheorem{theorem}{Theorem} 
\newtheorem{claim}{Claim}
\newtheorem{conjecture}{Conjecture} 
\newtheorem{corollary}{Corollary}
\newtheorem{proposition}{Proposition}
\newtheorem{lemma}{Lemma}

\newcommand{\bproof}{\setlength{\parindent}{0mm}{\bf Proof{~~}}}
\newcommand{\eproof}{$\Box$\setlength{\parindent}{5mm}} 

\DeclareMathOperator{\Int}{Int} 
\newcommand{\inter}[1]{\Int\,(#1)}

\DeclareMathOperator{\commonpast}{\downarrow\!\,}
\newcommand{\cpa}[1]{\commonpast#1}

\DeclareMathOperator{\commonfuture}{\uparrow\!\,}
\newcommand{\cfu}[1]{\commonfuture#1}

\DeclareMathOperator{\scommonpast}{\downarrow}
\newcommand{\scpa}[2]{\scommonpast_{#1}#2}

\DeclareMathOperator{\scommonfuture}{\uparrow}
\newcommand{\scfu}[2]{\scommonfuture_{#1}#2}

\newcommand{\blist}{\begin{list}{}{\setlength{\leftmargin}{4mm}
\setlength{\parindent}{0mm}\setlength{\parsep}{1mm}
\setlength{\topsep}{2mm}}}
\newcommand{\elist}{\end{list}}

\title{Morse index and causal continuity. 
A criterion for topology change in quantum gravity.}
\author{ H.F.Dowker${}^{a, 1,*}$, R.S.Garcia${}^{b,1,\dagger}$, 
         S.Surya${}^{c,2}$\\ 
        $\;$ \\ ${}^1$ Blackett Laboratory, Imperial College of
        Science Technology and Medicine,\\ London SW7 2BZ, United
        Kingdom \\ ${}^2$ TIFR, Homi Bhabha Rd, Mumbai 400 005,
        India.}

\begin{document}

\begin{titlepage}
\maketitle
\begin{abstract} 
\thispagestyle{empty} 
Studies in $1+1$ dimensions suggest that
causally discontinuous topology changing spacetimes are suppressed in 
quantum gravity. Borde and Sorkin have conjectured that causal
discontinuities are associated precisely with index $1$ or $\nmion$
Morse points in topology changing spacetimes  
built from Morse functions. We establish a weaker form of this
conjecture. Namely, if a Morse function $f$ on a compact cobordism has
critical points of index $1$ or $\nmion$, then all the Morse
geometries associated with $f$ are causally discontinuous, while if
$f$ has no critical points of index $1$ or $\nmion$, then there exist
associated Morse geometries which are causally continuous.

\end{abstract}
\vspace{0.1 cm}
\noindent ${}^a$dowker$@$ic.ac.uk, ${}^b$garciars$@$ic.ac.uk,
${}^c$ssurya$@$ tifr.res.in

\noindent ${}^*$ Present address: Dept. of Physics, QMW, London E1 4NS, U.K.

\noindent ${}^\dagger$ Present address: DAMTP, University of Cambridge,
Cambridge, CB3 9EW, U.K.

\end{titlepage}

\section{Introduction}

Within a gravitational Sum Over Histories (SOH), it is useful to obtain
selection rules to eliminate certain topological transitions, based on
general physical grounds, for example \cite{gibbons92}. A reasonable
criterion is that the quantum field propagation on the spacetime be finite,
and Sorkin has conjectured \cite{surya97,dowker97,borde99} that this can
only occur when the spacetime is causally continuous. It is therefore of
interest to determine whether a given topological transition admits a
causally continuous history or not. Guided by elements in surgery theory,
it was further conjectured by Borde and Sorkin \cite{borde99} that causal
continuity of a spacetime constructed from a Morse function 
depends on the indices of the critical points of
that Morse function . In this paper, we present a proof of a weaker form of
this second conjecture, which nevertheless can be used to determine an
appropriate selection rule for topology change.

There are no compelling reasons to exclude topology change from quantum
gravity. Indeed, our experience of the inconsistency of the relativistic
quantum mechanics of a fixed number of particles and the potential presence
in quantum gravity of states that correspond to particles (topological
geons) suggest that topology change {\it must} be included.  In the SOH
approach to quantum gravity
\cite{hawking78a,hawking78b,horowitz91,gibbons92,vilenkin94,carlip98}
topology change seems very naturally accommodated.  Moreover we can
reconcile topology change and causality, evading Geroch's theorem
\cite{geroch67} by letting the Lorentzian metric be degenerate at a finite
number of isolated singularities. This was the original motivation to
consider Morse spacetimes, which depart only mildly from being globally
Lorentzian, and can be defined in all possible cobordisms.

Let 
$(\cM,V_0,V_1)$ be a compact $n$-dimensional cobordism, 
with $\pd\cM=V_0\amalg V_1$, $h$ be a  Riemannian metric
on $\cM$ and $f:\cM\rar [0,1]$ a Morse function, with $f^{-1}(0)=V_0$,
$f^{-1}(1)=V_1$, which has $r$ critical points, $\{p_k\}$, in the
interior of $\cM$.  The critical points of 
$f$ are those where $\pd_\mu f=0\;\forall\mu$. What characterises a
Morse function is that the Hessian $\pd_{\mu}\pd_{\nu}f$ of $f$ is an
invertible matrix at each of its critical points, often
called Morse points. The Morse index $\la_k$ of the critical point
$p_k$ is the number of negative eigenvalues of the matrix
$(\pd_{\mu}\pd_{\nu}f)(p_k)$.  From $h$, $f$ and an arbitrary real
constant greater than $1$, denoted $\zeta$, we can define a {\it Morse
metric} in $\cM$:
\be\label{morsemetric.eq} 
g_{\mu\nu}=(h^{\rho\lambda}\pd_\rho f\pd_\lambda f)h_{\mu\nu} -\zeta
\pd_\mu f \pd_\nu f,
\ee 

This metric is Lorentzian everywhere, except at the critical points of
$f$. The {\it Morse geometry} $(M,g)$ associated with $\cM$, $h$, $f$
and $\zeta$ is the globally Lorentzian spacetime induced in
$M=\cM-\{p_k\}$, the manifold that remains after excising the critical
points from $\cM$.\footnote{The term ``geometry'' here does not 
imply quotienting by the action of any group of diffeomorphisms. 
We use it to distinguish the current case from future work in which the 
Morse point will regain its status as a physically present 
point and the pair $(\cM, g)$ will be known as 
a {\it Morse spacetime}.} We will adopt the notation that calligraphic
capital letters refer to the unpunctured manifolds and their Roman 
counterparts to the corresponding punctured manifolds.
  A general cobordism $\cM$ can be decomposed into a
finite sequence of elementary cobordisms $\cE_k$, each containing one
critical point. If $\cE$ is an elementary cobordism, we will refer to
a Morse geometry $(E,g)$, defined in $E=\cE-p$, as an elementary
Morse geometry. The vector field $h^{\mu\nu}\pd_\nu f$ is time-like
with respect to the metric $g$, so that we can think of $f$ as a
time-function on $(M,g)$. We refer the reader to
\cite{borde99,surya97,dowker97} for a more expository account.

	As detailed in \cite{surya97}, surgery theory suggests that
the index of $f$ at its critical points determines the ``continuity''
in the causal structure of a Morse geometry. For example, 
when the index is $1$ disconnected regions of space seem 
suddenly to come into contact with one another. 
Borde and Sorkin put forward the following conjecture:
\begin{conjecture}\label{bordesorkin.conjecture}
{\bf [Borde-Sorkin conjecture]} Given a compact cobordism $\cM$ and a
Morse function $f:\cM\rar\mathR$ with critical points $\{p_k\}$, then
a Morse geometry $(M,g)$ defined through $f$, is causally continuous
if and only if none of the points $p_k$ has Morse index $1$ or
$\nmion$.
\end{conjecture}
In \cite{borde99}, we studied a particular class of Morse 
geometries defined
in the neighbourhood of a critical point. From Morse theory we know
that around a critical point $p$ of a Morse function $f$ in an
$n$-dimensional cobordism, there is a round neighbourhood
$\cD_{\epsilon}$, of radius $\epsilon$, in local coordinates $(x_1,\cdots,
x_\lambda, y_1,\cdots y_{n-\lambda})$, in which the Morse function
takes the canonical form \cite{milnor65}:
\begin{equation}
\label{morselemma.eq} 
	f= c-\sum_{i=1}^\lambda x_i^2 + \sum_{j=1}^{n-\lambda} y_j^2.
\end{equation}
$c=f(p)$ is the critical value and $\lambda$ is the Morse index
of $p$. The neighbourhood Morse geometry we studied was 
defined on a particular neighbourhood of $p$ in $\cD_\epsilon$
(called $Q_\delta$ and chosen so that no spurious causal discontinuity 
could arise from the boundaries).  
 The metric was constructed from  
$f$ and the Cartesian flat Riemannian metric with
interval $ds_h^2=\sum_{i=1}^\lambda dx_i^2 + \sum_{j=1}^{n-\lambda}
dy_j^2$. 
We verified conjecture
\ref{bordesorkin.conjecture} for these ``Cartesian'' neighbourhood 
Morse geometries.

For a full proof of the conjecture, our earlier result needs to be
generalised in two main directions. First we have to verify
that the conjecture holds in neighbourhood Morse geometries 
constructed from arbitrary Riemannian metrics. Secondly we need
to embed the neighbourhood Morse geometries in the original Morse
geometry  
and show that the latter is causally continuous if and only if the 
neighbourhoods are. In this work we present progress made along these two
lines.

After a section containing background material, 
we start by investigating the general  Morse geometry  in the 
neighbourhood of a
Morse point $p$.  In section 3 we use results from dynamical systems
to establish
certain topological properties of the flow of the timelike vector field
$\xi^{\mu}=h^{\mu\nu}\pd_{\nu}f$. This helps us construct  past and future
sets that can be identified with the chronology of the critical point. 

In section 4 we show that that any Morse geometry 
$(M,g)$ constructed from a Morse function which  has an index $1$ or $\nmion$
point is causally discontinuous.

In section 5 we consider cobordisms $\cM$ with Morse functions $f$ with no index
$1$ and $n-1$ points. Here, we  show that one can always
construct a causally continuous Morse geometry built from $f$.  
We prove this in several steps.  First, we establish that a
Morse geometry $(M,g)$ is causally continuous if and only if all the
elementary Morse geometries which stacked together give $M$ are causally
continuous. Then we show that an elementary Morse geometry $(E,g)$
with a neighbourhood Morse geometry $(N,g)$ embedded in it, is causally
continuous if and only if $(N,g)$ is. Finally we show that we can always
choose an appropriate Riemannian metric which guarantees that 
there is a causally continuous neighbourhood Morse geometry around
each Morse point. 

In section 6 we summarise our results by stating them together
as a proposition which is a weaker version of the Borde-Sorkin
Conjecture.

\section{Review of Causal Structure}
\label{cs.section}

We review some relevant material on the causal structure of
globally Lorentzian spacetimes \cite{ellis73,sachs73,penrose72,beem81}. We
also include a few simple extensions of the usual results that will be
needed in later sections. All spacetimes in this paper are assumed to be
time-orientable and distinguishing and if a spacetime includes boundaries
they are restricted to be spacelike initial and final boundaries.

A timelike curve in a spacetime $(M,g)$ is a differentiable curve
$\gamma:\mathR\rar M$ whose tangent vector is everywhere timelike. A causal
curve is one whose tangent vector is timelike or null. We say that a
timelike(causal) curve starting at $x\in M$ is future-inextendible if it
has no future endpoint in $M$ except possibly on the final boundary of
$M$. One defines past-inextendible timelike(causal) curves similarly. Given
a pair of points $x$, $y$ in $M$, we write $x<<y$ if there exists a
future-directed timelike curve from $x$ to $y$, and $x<y$ if there exists a
future-directed causal curve from $x$ to $y$. The chronological future of a
point $x\in M$ is $I^+(x)= \{ y:\; x<<y\}$, while its causal future is
$J^+(x)= \{ y:\; x<y\}$. The chronological and causal pasts are defined
dually.

Clearly,  $I^{\pm}(x)\subset J^{\pm}(x)$. We also have (i)
$I^{\pm}(S)$ is open for every set $S\subset M$, (ii) $J^+(x)\subset
\ovl{I^+}(x)$ for every point in the interior of 
$M$, (iii) $x<y$ and $y<<z\; \Rar \; x<<y$ and (iv) $y\in
\ovl{I^+}(x)$ and $z\in\ovl{I^+}(y)\;\Rar
\;z\in\ovl{I^+}(x)$. 
Again, the dual statements hold.
For $U\subseteq M$, we write $I^+(x,U)$ to denote the
chronological future of $x$ in the spacetime $(U, g_{|U})$. The sets
$I^-(x,U)$ and $J^\pm(x,U)$ are defined similarly.  An open subset $U$
of a globally Lorentzian spacetime $(M,g)$ is said to be $I$-convex in
$M$ if for any pair of points $x$,$y\in U$ we have $I^+(x)\cap
I^-(y)\subset U$. It can be readily verified that if $U$ is
$I$-convex, then for any $x\in U$, $I^\pm(x,U)=I^\pm(x)\cap U$ and
that the interval $I^+(V)\cap I^-(W)$ between any two subsets $V$, $W$
of $M$ is $I$-convex.

The common past of an open set $S$ is $\cpa{S} \equiv \inter{\{q:\; q<<
s\;\forall\, s\in S\}}$, its common future $\cfu{S}$ is defined
dually.  It is immediate that $I^-(x) \subset \cpa{I^+(x)}$ and
$I^+(x) \subset\cfu{I^-(x)}$. If $U$ is a subspacetime of $M$ then
the common past relative to U of open set $S\subset U$ is written 
$\scpa{U}{S} \equiv \inter{\{q: q \in I^-(s, U),\ \forall s \in S\}}$.

\definition{A spacetime $(M,g)$ is causally continuous when either of the
following equivalent conditions holds:
\begin{list}{}{\setlength{\parindent}{0mm}\setlength{\leftmargin}{0mm}}
\item (A) $I^-(x)=\cpa{I^+(x)}$ and $I^+(x)=\cfu{I^-(x)}$ for every
point $x$ in $M\!-\!\pd M$.
\item (B) $x\in \ovl{I^-(y)}\, \Leftrightarrow\, y \in \ovl{I^+(x)}$ 
for every pair of points $x, y\in M$.
\end{list}\label{cc.definition}}

These are two of the six characterisations of causal continuity
given by Hawking and Sachs \cite{sachs73}, with a difference:  
Hawking and Sachs  
enforce (A) in the whole of $M$, which is assumed
boundaryless. A Morse geometry $M$ contains spacelike boundaries $V_0$ and
$V_1$, where the condition (A) trivially fails and hence the need for a
modification. Condition (B), however, can be extended to the
boundary without difficulty.

We also need the definitions  of causality, strong causality, stable
causality and global hyperbolicity. A spacetime $(M,g)$ is {\it
causal} if there are no closed causal curves in $M$. A spacetime
$(M,g)$ is {\it strongly causal} if every point in $M$ has
neighbourhoods that no causal curve intersects more than once. 
A spacetime $(M,g)$ is {\it stably causal} if
there exists a spacetime $(M,g')$ such that the lightcones of $g'$ are
everywhere wider than those of $g$ and which is causal. Stable
causality is equivalent \cite{ellis73} to the existence of a global
time function on $M$. A spacetime is {\it globally hyperbolic} if it
contains a spacelike hypersurface which every inextendible causal
curve intersects at exactly one point. The conditions listed are
arranged from stronger to weaker in the chain of implications: global
hyperbolicity $\Rar$ causal continuity $\Rar$ stable causality $\Rar$
strong causality $\Rar$ causality.

Given a closed set $V$ of a spacetime $(M,g)$ the {\it future domain
of dependence} of $V$, denoted $D^+(V)$, is the set of all points $x\in
M$ such that every past-inextendible causal curve through $x$
intersects $V$. Similarly one defines the past domain of dependence
$D^-(V)$. Their union, $D(V)=D^+(V)\cup D^-(V)$, is the {\it
domain of dependence} of $V$. It consists of all points $x\in M$ such
that every inextendible causal curve through $x$ intersects
$V$. Clearly $V\subset D(V)$. For any closed $V\subset M$, the set
$\inter{D(V)}$ is $I$-convex.
A set $V$ in $M$ is said to be
achronal if no two points in $V$ are chronologically related. It is
shown in~\cite{ellis73} that if $V$ is a closed achronal set in $M$
then $\inter{D(V)}$ is globally hyperbolic and therefore causally
continuous.

We finally note the special properties of Morse
geometries $(M,g)$ which, with their Morse points excised,  
are globally Lorentzian. Any Morse geometry $(M,g)$ is stably causal, since
the Morse function $f$ on $M$ is a global time function. In fact, causal
continuity is the weakest of the conditions listed above
which is not guaranteed hold in all Morse geometries, as we will see in
the next few sections. 

\section{Dynamical systems in a Morse neighbourhood}\label{topeq.section}

As mentioned in the introduction, in \cite{borde99} conjecture
\ref{bordesorkin.conjecture} was verified for spacetimes defined in a
certain punctured neighbourhood $Q$ of a critical point, with metric
constructed from the Cartesian Riemannian metric $\delta_{\mu\nu}$. These
neighbourhood Morse geometries are to be regarded as embedded in a Morse
geometry $(M,g)$. If we start from a general $(M,g)$, the Riemannian
metric $h$ from which $g$ is constructed will usually be different from
$\delta_{\mu\nu}$ in any neighbourhood of a Morse point. In this section we
investigate the causal structure in these more general neighbourhood
Morse geometries using some results from dynamical systems.

Consider the neighbourhood Morse geometry $(\De, g)$ around $p$ of index
$\lambda$ in which the Morse function takes the form
$f=c-\sum_1^{\lambda}(x^i)^2 +\sum_1^{\nmila}(y^j)^2$
and $\De = \{ (x^i, y^j): 
0<\sum_1^{\lambda}(x^i)^2 +\sum_1^{\nmila}(y^j)^2 <\epsilon^2\}$.
We place no
restrictions on the Riemannian metric $h_{\mu\nu}$ in equation 
(\ref{morsemetric.eq}).  In what follows it will sometimes be
convenient to group together the coordinates $(x^i,y^j)$ in a single set
$\{X^{\mu}\}$ with $X^{\mu}= x^{\mu}$ when $\mu\leq \lambda$ and $X^{\mu}=
y^{\mu-\lambda}$ when $\mu>\lambda$.

In our earlier investigations of neighbourhood Morse geometries with
$h_{\mu\nu}=\delta_{\mu\nu}$ \cite{borde99}, we defined certain past and
future sets, $\cP$ and $\cF$, which could be interpreted as the
chronological past and future  of the critical point. These played a crucial
role in the analysis, and so our first step is to construct $\cP$ and
$\cF$ for arbitrary $h_{\mu\nu}$.

\begin{figure}[ht]
\centering
\resizebox{!}{2.2in}{\includegraphics{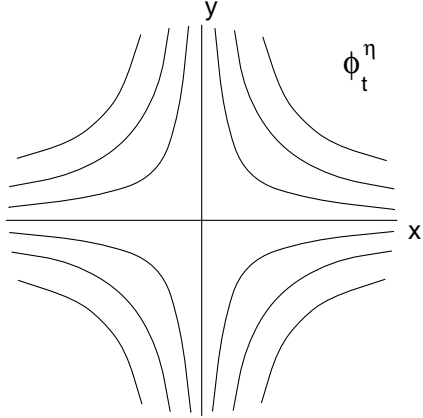}}
\caption{{\small Integral flow of the vector field $\eta = (-2x, 2y)$, 
the gradient-like vector field for the Morse function $f=-x^2+ y^2$ 
in the 2-dimensional disc.}\label{etaflow.fig}}
\end{figure}

We start by looking at the  vector field 
$\eta^\mu = \delta^{\mu\nu} \pd_{\nu}f$, which is a 
gradient-like vector field associated with $f$. This means that $\eta$ 
satisfies the conditions (i) $\eta^\mu$ is transverse to $f$, 
$\eta^\mu\pd_\mu f >0$, and (ii) in some neighbourhood of the Morse 
point , $\cDe$, $\eta^\mu$ takes
the canonical form,  $\eta^\mu=(-x_1 \cdots -x_{\lambda},y_1 \cdots
y_{n-\lambda})$, in local coordinates in which the Morse function 
takes {\it its} canonical form \cite{milnor65}. 
Figure~\ref{etaflow.fig}
shows some representative integral curves of $\eta$ in two 
dimensions, for an index $1$ critical point,  which are hyperbolae.  
Morse theory tells us, further,   that the integral
submanifolds of $\eta$ through $p$ (referred to as ``basins'' in the
language of dynamical systems)  are the discs $\cDela$  and $\cDenmila$
constructed from integral curves which end and begin at the critical
point, respectively, 
\bea \cDela &=&\left\{(x^1,\dots,x^{\lambda},0,\dots,0):\, 
\sum {x^i}^2 < \epsilon\right\}\nonumber\\
\cDenmila &=&\left\{(0,\dots, 0,y^1,\dots,y^{\nmila}):\, 
\sum {y^j}^2< \epsilon\right\}\nonumber
\eea

Consider also the vector field, $\xi^\mu=
h^{\mu\nu}\pd_\nu f$ for the neighbourhood Morse spacetime $(\De,g)$, which is
manifestly time-like with respect to $g$. 
When $g = \delta$, as in \cite{borde99}, $\xi=\eta$, but in general 
they are different and $\eta$ will not necessarily be timelike. 
Since $\xi$ is timelike, the properties of its flow are
important to the causal structure. We will now relate it to the 
flow of $\eta$ which we know exactly.
Indeed, we will show that 
the flows of $\xi$ and $\eta$ are topologically equivalent, in the
sense that the integral curves of one are mapped homeomorphically onto the
integral curves of the other. In particular, this implies that the basins
of $\xi$ swept out by the timelike integral curves are topologically
$D^{\la}$ and $D^{\nmila}$.  This is exactly what we need to construct
$\cP$ and $\cF$. 

We begin the proof of the topological equivalence of $\xi$ and $\eta$ by
first introducing certain properties of dynamical systems.

A  {\it dynamical system} in $\mathR^n$ is given by a
system of equations $\dot{X}^{\mu}=\xi^{\mu}(X)$, where $X^{\mu}$ are
the Cartesian coordinates in $\mathR^n$ and each $\xi^{\mu}$ a
differentiable real function on $\mathR^n$.  Solving  this system
amounts to finding its integral curves. That is,  for each point $X_0\in
\mathR^n$, we seek the unique curve $X(t)$ which satisfies, (i) $X(0) = X_0$
and (ii) $ \dot{X}^{\mu}(t) = \xi^{\mu}(X(t))$.  

The {\it flow} of $\xi$, which we denote by $\phi_t^{\xi}$, is the one
parameter family of local diffeomorphisms defined by pushing points
along the integral curves of $\xi$. More specifically, in a
neighbourhood of each point $x$ and for each $t$ in some finite range
$(-\epsilon, \epsilon)$, the map $y\rar \phi_t^{\xi}(y)$ is a
diffeomorphism. A point $p$ is a fixed point of the dynamical system
if $\xi^{\mu}(p)=0$ for every $\mu$. 

A {\it linear system}, with fixed point at the origin, is defined by the
equations, $\dot{X}^{\mu}=A^{\mu}_{\,\nu}\, X^{\nu}$, where
$A^{\mu}_{\,\nu}$ is some constant matrix. An arbitrary dynamical system
$\xi^{\mu}$ admits a linear approximation around an isolated fixed point
$p$ which we denote $\tilde{\xi}^{\mu}$ and is given by $\dot{X}^{\mu}=
\tilde{\xi}^{\mu}=\left(\pd_{\nu}\,\xi^{\mu}\right)(p)X^{\nu}$.

The integral flows $\phi^{\xi}_t$ and $\phi^{\eta}_t$ of two vector fields
$\xi$, $\eta$ in $\mathR^n$ are said to be {\it topologically equivalent}
if there exists a bijection $\Psi: \mathR^n\rightarrow \mathR^n$ taking the
flow $\phi_t^{\xi}$ to the flow $\phi_t^{\eta}$, so that
$\Psi\circ\phi_t^{\xi}= \phi_t^{\eta}\circ \Psi$ for any $t\in \mathR$.
The two topologically equivalent flows are thus  related by the change of
coordinates $\Psi$.  

We now state two important theorems on dynamical systems, given for example in
\cite{arnold92,perko91}.

\begin{theorem}
\label{llte.theorem} Two linear systems having no eigenvalues with real 
part zero are topologically equivalent if and only if the number of
eigenvalues with negative real part is the same for the two systems.
\end{theorem}

\begin{theorem}\label{nllte.theorem}
Let $\phi_t^{\xi}$ be the integral flow of a dynamical system
 $\xi^{\mu}$ in $\mathR^n$ which has a fixed point at the origin
 $\vec{0}$ and $\phi_t^{\tilde{\xi}}$ the integral flow of its linear
 approximation. Suppose the matrix $A^{\mu}_{\,\nu}=
\pd_{\nu}\,\xi^{\mu}(\vec{0})$ which determines the linear system
$\tilde{\xi}$ has no purely imaginary eigenvalues. Then there is a
neighbourhood $U$ of the origin and a homeomorphism $\Psi$ from $U$
onto another neighbourhood $V$ of the origin such that $\Psi\circ
\phi_t^{\tilde{\xi}} = \phi_t^{\xi}\circ \Psi$.
\end{theorem}

What this theorem tells us is that a non-linear system is
topologically equivalent to its linearisation in a neighbourhood of
the fixed point, provided the linear system has no purely imaginary
eigenvalues. Clearly, the range of $t$ involved in $\Psi\circ
\phi_t^{\tilde{\xi}} = \phi_t^{\xi}\circ \Psi$ is restricted for each
$q\in U$ so that $\phi_t^{\tilde{\xi}}(q)$ lies in $U$. Using  these
results we can now establish,

\begin{lemma}\label{topeq.lemma} Let $\cDe$ be the round neighbourhood 
of radius $\epsilon$ about
the critical point of the Morse function $f$ and $g$ be the Morse metric
(\ref{morsemetric.eq}) constructed from $f$ and a Riemannian metric
$h$. Let $\phi_t^{\eta}$ be the flow of the gradient-like vector field
$\eta^{\mu}=\delta^{\mu\nu}\pd_{\nu}f$ in $\cDe$ and $\phi_t^{\xi}$ that of
the vector field $\xi^{\mu}=h^{\mu\nu}\pd_{\nu}f$ in $\cDe$, which is
timelike with respect to $g$.  Then there is a neighbourhood $\cDeprime$
with $\epsilon' < \epsilon$ and a homeomorphism $\Psi: \cDeprime \rar
\cN\subset\cDe$, which is a topological equivalence between $\phi_t^{\eta}$
in $\cDeprime$ and $\phi_t^{\xi}$ in $\cN \equiv \Psi(\cDeprime )$.
\end{lemma}

\bproof As we observed earlier, both dynamical systems have a single fixed
point at $p$, which we take to be the origin in the coordinates
$\{X^{\mu}\}$. Since the disc is homeomorphic to $\mathR^n$, theorems
\ref{llte.theorem} and \ref{nllte.theorem} can be readily applied.

The flow $\phi_t^\eta$ is associated with the linear dynamical system
$\dot{X}^{\mu}=\eta^{\mu}= 2\Lambda^{\mu}_{\,\nu}X^{\nu}$, where
$\Lambda$ is the diagonal matrix $(-1,\dots,-1,1,\dots,1)$, with
$\lambda$ negative eigenvalues and $\nmila$ positive
eigenvalues\footnote{We regard $\Lambda^{\mu}_{\,\nu}$ as the result
of raising the Hessian $\Lambda_{\mu\nu}={\frac{1}{2}}\pd_{\mu}\pd_{\nu}f$ 
with the
Cartesian flat metric $\delta^{\mu\nu}$. We raise and lower the
indices on $\Lambda$ with $\delta$ so that its entries have always the
same value, irrespective of the index position.}. The flow of
$\phi_t^\xi$ is determined by the dynamical system
$\dot{X}^{\mu}=\xi^{\mu}= h^{\mu\nu}\pd_{\nu}f$, which is in general
non-linear, since the metric components $h^{\mu\nu}$ are arbitrary
functions of the coordinates $\{X^{\mu}\}$. The linearisation of this
dynamical system around its fixed point $p$ is given by:
\be
\dot{X}^{\mu}= \pd_{\rho}(h^{\mu\nu}\pd_{\nu}f)\, X^{\rho}
\ee
where the partial derivatives are evaluated at $p$. Since here
$\pd_{\mu}f=0$, the derivatives of $h^{\mu\nu}$ do not appear
and we are left with:
\bea\label{linearisation.eq}
\dot{X}^{\mu}&=& H^{\mu\nu}(\pd_{\rho}\pd_{\nu}f)_p\,
X^\rho\nonumber \\
&=& 2 H^{\mu\nu} \Lambda_{\nu\rho}\, X^\rho
\eea
where $H$ is the symmetric positive-definite matrix $H^{\mu\nu}=
h^{\mu\nu}(p)$. Denoting  the matrix that governs the linear 
system~(\ref{linearisation.eq}) by $A^{\mu}_{\,\nu}= 2 H^{\mu\rho}
\Lambda_{\rho\nu}$, we may now assert:

\begin{claim}\label{eigenvalues.claim} The eigenvalues of 
$A^{\mu}_{\nu}$ are all real. There are $\lambda$ negative ones and
$\nmila$ positive ones.
\end{claim}
\bproof
By symmetry and positive-definiteness, there is an orthogonal
transformation that takes $H$ into a diagonal matrix $D$ with strictly
positive entries. That is $H=ODO^T$. The matrix $H^{1/2}=OD^{1/2}O^T$
is also symmetric and positive-definite. We know that a similarity
transformation preserves eigenvalues, while conjugacy\footnote{The
conjugate of a matrix $A$ by an invertible matrix $B$ is the matrix
$BAB^T$.}  preserves the number of negative, positive and zero
eigenvalues (this is Sylvester's theorem). Now:
\newcounter{temp}
\setcounter{temp}{\value{equation}}
\setcounter{equation}{0}
\renewcommand{\theequation}{\roman{equation}}
\bea
H^{-1/2} A H^{1/2} &=& 2 H^{1/2} \Lambda H^{1/2}\qquad \mbox{and}\\
H^{-1/2} (2 H^{1/2} \Lambda H^{1/2}) (H^{-1/2})^T &=& 2 \Lambda
\eea
\renewcommand{\theequation}{\arabic{equation}}
\setcounter{equation}{\value{temp}}
From equation (i) we see that the eigenvalues of $A$ are the same as
those of the symmetric matrix $2 H^{1/2} \Lambda H^{1/2}$, and hence they
must be real. Then equation (ii) tells us that the distribution of
eigenvalues of $A$ into positive, negative and zero is the same as for
the matrix $2 \Lambda$. Hence, the claim. \eproof

\begin{figure}
\centering
\resizebox{10cm}{!}{\includegraphics{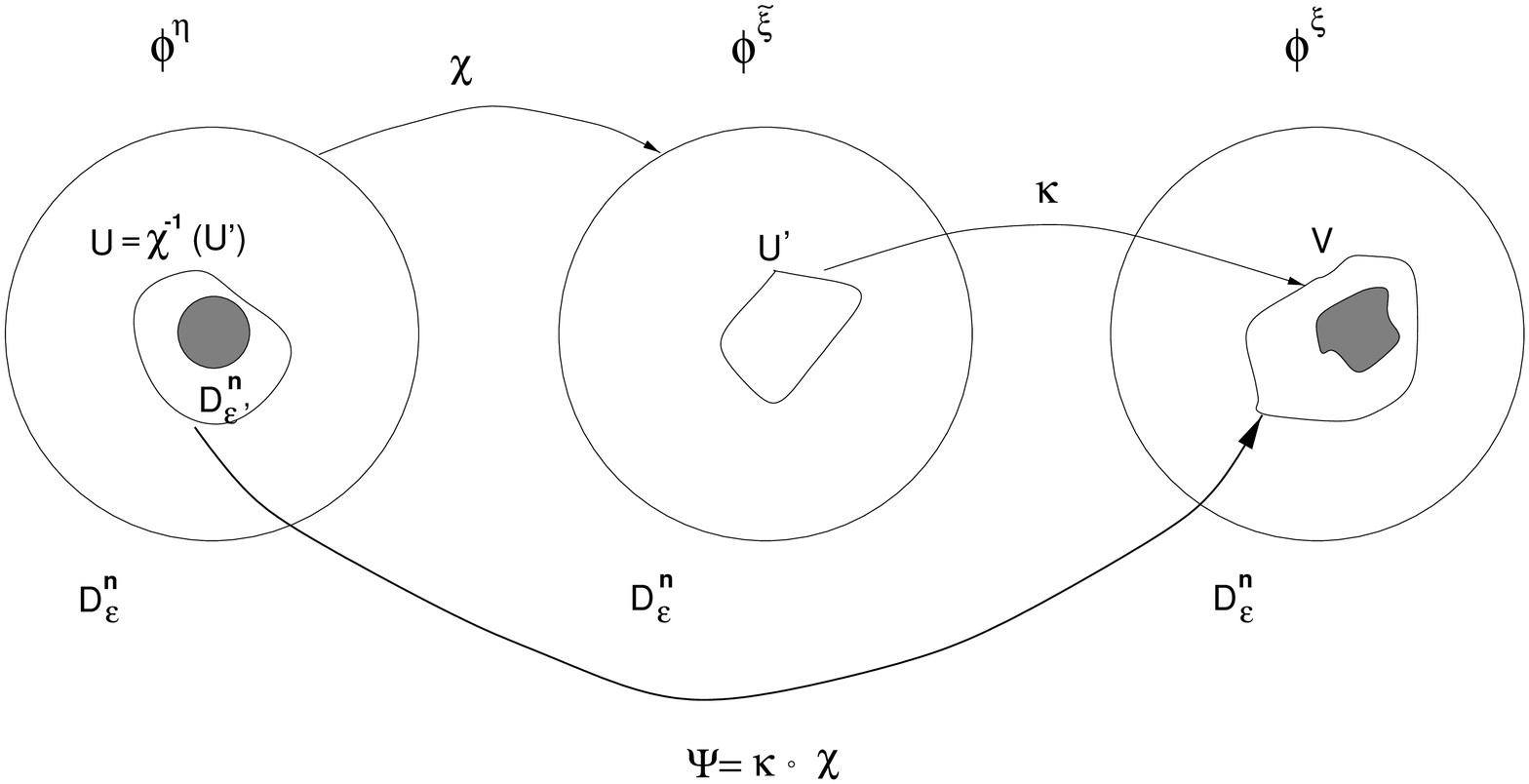}}
\vspace{4mm}
\caption{\small{We compare the flows of 
the gradient-like vector field $\eta^{\mu}$ and the timelike vector
field $\xi^{\mu}=h^{\mu\nu}\pd_{\nu}f$ using the linearisation
$\tilde{\xi}^{\mu}$ of $\xi^{\mu}$ as an intermediate step. Here the
homeomorphism $\chi: \cDe\rar \cDe$ 
is a topological equivalence between
the linear flows of $\eta$ and $\tilde{\xi}$. The homeomorphism
$\kappa: U'\rar V$ gives a topological equivalence between the flows of
$\tilde{\xi}$ in $U'$ and $\xi$ in $V$. We will consider the
restriction of $\Psi= \kappa\circ\chi$ from $U=\xi^-1(U')$ to $V$, to a round
neighbourhood $\cDeprime$ of $p$, where the topological properties of the
flow of $\eta$ are easy to see.}\label{topeq.fig}}
\end{figure}

By claim~\ref{eigenvalues.claim} and theorem~\ref{llte.theorem} we see
that the flows associated with the linear systems $\eta$ and
$\tilde{\xi}$ are topologically equivalent in $\cDe$.
On the other hand, combining the claim with
theorem~\ref{nllte.theorem} we deduce that the flows of $\tilde{\xi}$
and $\xi$ are topologically equivalent in a neighbourhood of the
critical point.  We can then compose the corresponding homeomorphisms
to conclude that there is a neighbourhood $\cU$ of $p$ in $\cDe$ and a
homeomorphism $\Psi$ from $\cU$ onto another neighbourhood $\cV$ of
$p$ in $\cDe$ such that for any $q\in \cU$ and $t$ with
$\phi_t^{\eta}(q)\in \cU$ we have $\Psi\circ\phi_t^{\eta}(q)
=\phi_t^{\xi}\circ \Psi (q)$.  In figure~\ref{topeq.fig} we have
represented the two step construction of $\Psi$.

 The result now follows, for $\cU$ contains a neighbourhood
$\cDeprime$ of $p$, for some $\epsilon' < \epsilon$, and the restriction
$\Psi_{|\cDeprime}$ maps the flow of $\eta$ homeomorphically to the
flow of $\xi$ in $\cN \equiv
\Psi(\cDeprime)$. \eproof

As we stated earlier, our aim in proving the  topological equivalence
between $\xi$ and $\eta$ is that this implies that the basins of $\xi$ in
$\cN$ are topologically the same as those of $\eta$ in $\cDeprime$. We denote the
``ingoing'' one by  $\cNla\equiv\Psi (\cDeprimela)$ and the ``outgoing'' one
by $\cNnmila\equiv\Psi(\cDeprimenmila)$. 

Now, let us consider the neighbourhood Morse geometry $(N,g)$, where
$N=\cN\mip$. This is a subspace  of the original Morse geometry  
$(M,g)$. In $(N,g)$ the basins of $\xi$ are replaced by the two 
punctured disks, $\Nla=\cNla\mip$ and $\Nnmila= \cNnmila\mip$, which  are
swept out by congruences of timelike curves, that respectively ``end'' and
``begin'' at $p$.  Let us define the past set, $\cP\equiv I^-(\Nla,N)$ and 
the future set $\cF\equiv I^+(\Nnmila,N)$ from these punctured discs.   
Clearly, $\cP$ lies in the subcritical region of $N$ and $\cF$ in the
supercritical region. We now show that every point in $\cP$ is to the past
of every other point in $\cF$. 


\begin{claim}\label{PinpastofF.claim} Any pair of points $s$, $q$ 
with $s\in \cP$ and $q\in \cF$ satisfy $s\in I^-(q,N)$.
\end{claim}
\bproof 
By construction, for every $s \in \cP$  there exists an $s'\in \Nla $ 
such that $s\in I^-(s',N)$ and similarly, for every $q\in \cF$, there
exists a $q'\in \Nnmila$ such that $q\in I^+(q',N)$. 
Consider  open neighbourhoods $U_{s'}$ of $s'$ and
$U_{q'}$ of $q'$ with $U_{s'}\subset I^+(s,N)$ and $U_{q'}\subset
I^-(q,N)$. We now show that there are points $u\in U_{s'}$ and $v\in
U_{q'}$ connected by an integral curve of the timelike vector field
$\xi$, contained in $N$, which implies that $u\in I^-(v,N)$ as required. 

The flow of $\xi$ in $N$ is mapped to the flow of $\eta$ in $\Deprime 
\equiv \cDeprime\mip$
by $\Psi^{-1}$. So, consider $\Psi^{-1}(s') \in \Deprimela \equiv
\cDeprimela\mip$ with
coordinates $(\vec{x}_{s'}, \vec{0})$ and $\Psi^{-1}(q') \in
\Deprimenmila \equiv
\cDeprimenmila\mip$ 
with coordinates $(\vec{0}, \vec{y}_{q'})$. Consider a
point close to $(\vec{x}_{s'}, \vec{0})$, i.e., $(\vec{x}_{s'}, \delta
\vec{y}_{q'})$, where $\delta>0$ 
is some (small) constant.  The integral
curve of $\eta$ through $(\vec{x}_{s'}, \delta \vec{y}_{q'})$ is given by
$(e^{-2t} \vec{x}_{s'}, e^{2t} \delta \vec{y}_{q'})$.  This curve passes
through $(\delta \vec{x}_{s'}, \vec{y}_{q'})$ when $t = {\frac{1} {2}}\ln
{\frac{1}{\delta}}$.  The image of this curve under $\Psi$ in $N$ is a
timelike curve and by continuity $\delta$ can be chosen small enough so
that it passes arbitrarily close to $s'$ and $q'$. \eproof

We end this section with a couple of results for arbitrary $\lambda$. 
These do not use the dynamical systems technology, but will be useful in
the proofs of the next few sections.  

First we show that no causal curve that begins and ends close enough to $p$
can stray too far from it.

\begin{claim}\label{curvesconfined.claim} Let $(U,g)$ be a
neighbourhood Morse geometry around a Morse point $p$ of the Morse 
geometry 
$(M,g)$. Then there is a neighbourhood $U'\subset U$ of $p$ such that
any causal curve $\gamma$ between two points $x$, $y$ in $U'$ must be
contained in $U$.
\end{claim}

\bproof 
Consider $x,y \in U$ and $\gamma$ a future directed causal curve from 
$x$ to $y$. Suppose that $\gamma$ leaves $U$ between $x$ and $y$.
 Since $\gamma$ is causal we must have
$g_{\mu\nu}dx^{\mu}dx^{\nu} \le 0$ at every point in $\gamma$. For the
Morse metric $g_{\mu\nu}$ this means: 
\[
h^{\rho\sigma}\pd_{\rho} f\,\pd_{\sigma}f\, 
h_{\mu\nu}dx^{\mu}dx^{\nu}- \zeta\,\pd_{\mu}f\,\pd_{\nu}f\,
dx^{\mu}dx^{\nu}\le0
\]
\[\hspace{-20mm}\mbox{or}\quad\;\zeta\,(\pd_{\mu}f\, dx^{\mu})^2 \ge
h^{\rho\sigma}\pd_{\rho} f\,\pd_{\sigma}f\, h_{\mu\nu}dx^{\mu}dx^{\nu}
\]
Since $\gamma$ is future-directed, $\pd_{\mu}f\,
dx^{\mu}>0$ at every point of $\gamma$ so that taking square  
roots and then rearranging we get: 
\[
\zeta^{1/2}\pd_{\mu}f\, dx^{\mu}
-(h^{\rho\sigma}\pd_{\rho} f\,\pd_{\sigma}f\,)^{1/2}\, 
(h_{\mu\nu}dx^{\mu}dx^{\nu})^{1/2}\ge 0
\]
Integrating this expression along $\gamma$
we obtain:
\be\label{df-int.eq}
\zeta^{1/2}\left(f(y)-f(x)\right)-
\int_{x}^{y}(h^{\rho\sigma}\pd_{\rho} f\,\pd_{\sigma}f\,)^{1/2}
(h_{\mu\nu}dx^{\mu}dx^{\nu})^{1/2} \ge 0
\ee
We now establish a lower bound  for the above integral. Let
$D_A\subset D_B\subset U$ be round, open $n$-discs 
centred on $p$. Their boundaries are the $\nmion$-spheres $S_A$ and
$S_B$, respectively. Suppose $x, y\in D_A$, then the curve $\gamma$ must have
at least one arc $\gamma([t_A,t_B])$ with $t_x< t_A <t_B <t_y$ which lies
entirely in the cylinder $C=\ovl{D_B}-D_A$ and with
$x_A=\gamma(t_A)\in S_A$ and $x_B=\gamma(t_B)\in S_B$ and another 
such arc coming back in from $S_B$ to $S_A$. Since
$\pd_{\mu}f$ is nowhere vanishing in the compact set $C$, the function
$(h^{\mu\nu}\pd_{\mu}f
\pd_{\nu}f)^{1/2}$ must have a minimum $\mu>0$ in this region.  It follows
that, 
\be
\cint_{x_A}^{x_B}(h^{\rho\sigma}\pd_{\rho} f\,\pd_{\sigma}f\,)^{1/2}
(h_{\mu\nu}dx^{\mu}dx^{\nu})^{1/2}\geq \mu 
\cint_{x_A}^{x_B}(h_{\mu\nu}dx^{\mu}dx^{\nu})^{1/2}\nonumber 
\ee
 Finally, we must show that the Riemannian length of any path from
$S_A$ to $S_B$ is larger than some fixed positive number. Now for 
any Riemannian metric $h$ in a manifold $M$, the distance function
$d:M\times M\rar [0,\infty)$ defined by
\be
d(x,y)= Inf_{{}_{\gamma\in \Omega(x,y)}}\{\cint_{\gamma}
(h_{\mu\nu}dx^{\mu}dx^{\nu})^{1/2}\}
\ee
where $\Omega(x,y)$ is the set of piecewise differentiable curves from
$x$ to $y$, is continuous and satisfies $d(x,y)=0\;\Leftrightarrow\;
x=y$. It follows that for any two disjoint compact subsets $A$ and $B$
in $M$ the number $d(A,B)= Inf\,\{\,d(x,y): x\in A ,\, y\in B\}$ is
positive.
Hence, since $S_A$ and $S_B$ are disjoint
and compact, $d(S_A,S_B)$ is some positive number, 
$d_{AB}$. Thus
\be
\int_x^{y}(h^{\rho\sigma}\pd_{\rho} f\,\pd_{\sigma}f\, )^{1/2}
(h_{\mu\nu}dx^{\mu}dx^{\nu})^{1/2} \geq 2 \mu\, d_{AB}
\ee
which we can then insert in equation (\ref{df-int.eq}) to
obtain:
\be
f(y)- f(x) \ge 2 \zeta^{-1/2}\, \mu\, d_{AB}
\ee
Let $\alpha= \zeta^{-1/2}\,\mu\, d_{AB}$. 
If we define $U'=D_A\cap f^{-1}((c-\alpha,c+\alpha))$ then any causal
curve $\gamma$ between two points in $U'$ cannot leave $U$.\eproof

Finally we prove the almost obvious result:

\begin{claim}\label{dPdFaroundp.claim} Given a Morse geometry $(M,g)$,
let $N=\cN\mip$ be the punctured neighbourhood of a critical point
$p$ with $\cN$ as in lemma~\ref{topeq.lemma} so that
$\cP=I^-(\Nla,N)$ and $\cF=I^+(\Nnmila,N)$ are defined as before. Then any
neighbourhood $U$ of $p$ contains points in $\pd \cP$ and $\pd \cF$.
\end{claim}

\bproof Let $\tilde U$ be the connected component of 
$U\cap N$ around the critical point. Pick a point $x$ in $V_c\cap
\tilde U$ where $V_c \equiv \{x \in M: f(x) = c\}$ is the 
critical surface. The same arguments used in the proof of
claim~\ref{curvesconfined.claim} guarantee that there exists a
neighbourhood $U_x\subset \tilde U$ of $x$ with $U_x\cap \cF
=\emptyset$. Take any curve $\gamma: [0,1]\rar U'$ with $\gamma(0)= x$
and $\gamma(1)= y\in D^{\nmila}_{\epsilon}$ and let $\tau =
Sup_{\,t}\,\{\forall\, t'<t \; \gamma(t') \notin
\cF\}$. Then $\tau\in(0,1)$, since there is a neighbourhood of $y$ in
$\cF$, and $\gamma(\tau)\in \pd\cF$. That $U$ contains points in 
$\partial\cP$ is proved similarly. \eproof

\section{Index $\lambda = 1, \nmion$}\label{la1nmion.section}

We can now prove one half of the Borde-Sorkin conjecture.

\begin{lemma}\label{indexone.lemma} A Morse geometry $(M,g)$ 
with an index $1$ or $n-1$ Morse point $p$ is causally 
discontinuous. 
\end{lemma}

\bproof We consider only the index $1$ case, since the $\nmion$ result 
follows by the dual argument. Pick a neighbourhood $\cN$ of $p$ as in 
lemma \ref{topeq.lemma}, with coordinates $\{x,y^{\mu}\;\}$ so that
$f=c-x^2+\sum_{\mu=1}^{\nmion } (y^{\mu})^2$. Let $\Pi =\{q\in \cN: f(q)<c\}$
and $\Phi= \{q\in \cN : f(q)>c\}$ be the subcritical and supercritical regions
of the neighbourhood, respectively. $\Pi$ comprises 
two disconnected components, $\Pi_1$ and $\Pi_2$, corresponding to $x>0$ and
$x<0$, while $\Phi$ is connected. We know from lemma~\ref{topeq.lemma} that
the basins of the timelike vector field $\xi$ in $\cN$ have the same
topology and relative position as those of the gradient-like vector field
$\eta$ in $\Deprime$.  Thus, the outgoing basin is an $\nmion$-disc
$\cN^{\nmion}$ which being of codimension $1$, separates $\cN$ into two
pieces, one
containing $\Pi_1$ and the other $\Pi_2$.  The ingoing basin is a $1$-disc
$\cN^1$ which itself is divided in two by $\cN^{\nmion}$ so that it
stretches into $\Pi_1$ and $\Pi_2$ across the critical point 


In the associated Morse geometry $(N,g)$, where $N=\cN\mip$, define $\cP
\equiv I^-(N^1,N)$ and $\cF \equiv I^+(N^{\nmion},N)$ as before.  $\cP$
splits into $\cP_1\amalg\cP_2$, where $\cP_i= \cP \cap \Pi_i$.  Similarly
$\pd \cF$ has two components $\pd \cF_1$ and $\pd \cF_2$, one within each
of the halves in which $N^{\nmion}$ divides $N$. Choose $\pd \cF_1$ to be
in the half which also contains $\cP_1$.

Take points $s\in\cP_2 $ and $q\in\pd\cF_1$ close enough to $p$ so
that a timelike curve from $s$ to $q$, if there is one, must be
totally contained in $N$.  Note that claims
\ref{curvesconfined.claim} and \ref{dPdFaroundp.claim} guarantee the
existence of such $s$ and $q$ in $N$.  Then $s\notin I^-(q)$, since if
there were a timelike curve from $s$ to $q$ it would have to cross the
separating disc $N^{\nmion}$ and we would conclude that $q\in\cF$,
which is a contradiction, since $\cF$ is open. Moreover, since $\cF$
is a future set \cite{penrose72}, any point $q\in \pd \cF$ must
satisfy $I^+(q,N)\subset\cF$, which combined with
claim~\ref{PinpastofF.claim} implies $\cP\subset\scpa{N}{\cF}\subset
\scpa{N}{I^+(q,N)}$. Clearly, $\scpa{N}{I^+(q,N)}\subset
\cpa{I^+(q,N)}$, and since any $q'\in I^+(q)$ is in the chronological
future of some $q''\in I^+(q,N)$, we also have
$\cpa{I^+(q,N)}=\cpa{I^+(q)}$. It follows that $\cP\subset
\cpa{I^+(q)}$ and therefore $s\in\cpa{I^+(q)}- I^-(q)$. \eproof

Having dealt with this most general proof of causal discontinuity for all
index $1$ and $n-1$  neighbourhood Morse geometries, we would like 
to construct a similar proof of causal continuity for index $\neq 1,
n-1$. However, such a proof requires more than the topological equivalence
of $\xi$ and $\eta$, since we need not one family, $\xi$, 
but {\it all} possible families of timelike curves. 

However we do have examples of  causally continuous index $\neq 1, n-1$
neighbourhood  geometries from \cite{borde99}, 
namely the ``Cartesian'' ones.
We use this result crucially in the construction of
the main proposition. We now proceed to examine cobordisms with no index $1,
\nmion$ points.

\section{Cobordisms with no index $1$ or $n-1$ points}
\label{laneq1nmion.section}

A corollary of Lemma~\ref{indexone.lemma} 
is that a cobordism for which every Morse
function contains Morse points of index $1$ or $\nmion$ supports no
causally continuous Morse geometries.  We now show that any cobordism
which admits Morse functions without index $1$ or $\nmion$ points supports
causally continuous Morse geometries.  We will do so by combining the causal
continuity of the Cartesian neighbourhood Morse geometries, $(\Qd,g)$ of type
$(\la,\nmila)$ when $\la\neq 1,\nmion$ (see~\cite{borde99}), with the
``stacking'' and ``insertion'' results proved below. In the proofs we will 
be using the following two claims.

Let $U$ be an open subset of the spacetime $M$. We
have already seen that $\scpa{U}{I^+(y,U)}\subset\cpa{I^+(y)}\cap U$
for every $y\in U$. For a certain class of subsets $U$ the converse
also holds.

\begin{claim}\label{IUcap.claim} Let $(M,g)$ be an arbitrary
spacetime and $U$ an open subset of $M$ such that for every point
$x\in U$ we have $I^{\pm}(x,U)= I^{\pm}(x)\cap U$. Then for each
$y\in U$ we have $\scpa{U}{I^+(y,U)}= \cpa{I^+(y)}\cap U$.
\end{claim}

\bproof Suppose $x$ lies in $\cpa{I^+(y)}\cap U$. It must have a 
neighbourhood $U_x$ in $\cpa{I^+(y)}\cap U$. 
For each $y'\in I^+(y,U)$,  $U_x\subset I^-(y')\cap U =
I^-(y',U)$, so that $U_x\subset \scpa{U}{I^+(y,U)}$. \eproof  

The dual result is proved similarly.  Claim \ref{IUcap.claim} clearly
holds for any open subset $U$ which is $I$-convex in $M$. A related
result, important to our study of causal continuity is:

\begin{claim}\label{IUcapcd.claim} Let $(M,g)$ be an arbitrary
spacetime and $U$ an open subset of $M$ such that for every point
$x\in U$, $I^{\pm}(x,U)= I^{\pm}(x)\cap U$. If  $U$ is
causally discontinuous then $M$ is causally discontinuous.
\end{claim}
\bproof Suppose that for points $x$, $y$ in $U$ we have $x\in
\scpa{U}{I^+(y,U)}- I^-(y,U)$. We know that $x\in 
\cpa{I^+(y)}$. And since 
$I^-(y,U)= I^-(y)\cap U$, we must have $x\notin I^-(y)$. If 
$\scfu{U}{I^-(y,U)} \ne I^+(y,U)$ one proceeds similarly to show that 
$\cfu{I^-(y)\neq I^+(y)}$. \eproof

\subsection{Stacking cobordisms}\label{stack.section} 

\begin{lemma}\label{stacking.lemma} Consider a Morse geometry 
 $(M,g)$ defined through the Morse function $f$ and a Riemannian metric
$h$. Let $M_1= f^{-1}([0, b))$ and $M_2= f^{-1}((a,1])$, with $a<b$, so
that $M=M_1\cup M_2$. Then $(M,g)$ is causally continuous iff both
$(M_1,g)$ and $(M_2,g)$ are causally continuous.
\end{lemma}
\bproof Clearly $M_1$ and $M_2$ are $I$-convex in $M$, since any timelike
curve $\gamma$ between two points $x$ and $y$, in say $M_1$, must
satisfy $f(x)\leq f(\gamma(t)) \leq f(y)$ and hence be contained in
$M_1$.  Therefore, if either $M_1$ or $M_2$ is causally discontinuous,
then by claim \ref{IUcapcd.claim}, $M$ too will be causally
discontinuous.

Now, let both $M_1$ and $M_2$ be causally continuous.
Clearly, for $x\in M_1$,  $I^-(x)= I^-(x,M_1)$, so that
$\cpa{I^+(x)}\subset \cpa{I^+(x,M_1)} = I^-(x)$.  Similarly, for $y$
in $M_2$ we have $\cfu{I^-(y)}= I^+(y)$.  Thus, it remains for us to 
show  that $I^-(y)=\cpa{I^+(y)}$ for every $y\in M_2$. By a dual argument,
one would find that $I^+(x)=\cfu{I^-(x)}$ for every $x\in M_1$. 

Let us assume the contrary, i.e., $\exists \, y$ such that
$\cpa{I^+(y)}-I^-(y)\neq\emptyset$. This means that 
$\cpa{I^+(y)}-\ovl{I^-(y)} \neq \emptyset$. Let
$x\in\cpa{I^+(y)}-\ovl{I^-(y)} $. Now, $x$ must belong to $M_1-M_2\cap
M_1$. Otherwise, from convexity of $M_2$ and claim~\ref{IUcap.claim} we
would have $x\in\scpa{M_2}{I^+(y,M_2)}- I^-(y,M_2)$, thus contradicting the
causal continuity of $(M_2,g)$.

Next, consider a sequence of points $y_k \rightarrow y$ with $y_k \in
I^+(y)$. There exists a sequence of timelike curves $\gamma_k$ from
$x$ to $y_k$. Choose a number $d$ with $a< d < b$ so that 
$\Sigma \equiv f^{-1}(d) \subset M_1\cap M_2$ is a regular level surface of
$f$ and therefore a closed $\nmion$ manifold. Each $\gamma_k$ intersects
$\Sigma$ at a point $z_k$. Since $\Sigma$ is compact, a subsequence
of the $z_k$ converges to a point $z\in \Sigma$. For every $k$ such
that $z_k$ doesn't belong to this subsequence throw away $z_k$,
$\gamma_k$ and $y_k$ and relabel the remaining subsequences as $z_k$,
$\gamma_k$ and $y_k$.  Clearly $z\in \ovl{I^+(x)}$ which, together
with causal continuity of $M_1$ implies $x\in \ovl{I^-(z)}$.

Now, for any $z'\in I^-(z)$ there is a tail of the sequence $z_k$ 
contained in $I^+(z')$ so that $y\in \ovl{I^+(z')}$. Therefore, by
causal continuity of $M_2$ we have $z'\in\ovl{I^-(y)}$. Since
$I^-(z)$ contains points arbitrarily close to $z$, we see that
$z\in \ovl{I^-(y)}$. Transitivity of $\ovl{I^-}$ then implies 
that $x\in \ovl{I^-(y)}$,  a contradiction. \eproof 

\vspace{0.5cm} 
\begin{figure}[ht]
\centering
\resizebox{3.4in}{!}{\includegraphics{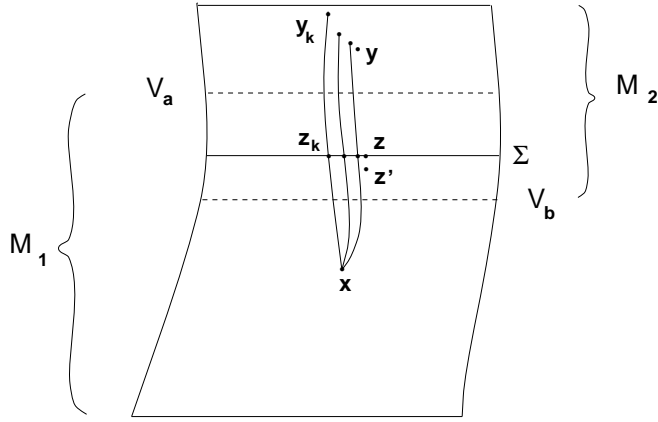}}
\vspace{0.5cm}
\caption{\small{Divide the Morse geometry $M$ in two
blocks $M=M_1\cup M_2$, bounded by level surfaces of the Morse
function $f$. If $(M_1,g)$ and $(M_2,g)$ are causally continuous,
so must be $M$.}\label{stack.fig}}
\end{figure}

Any Morse geometry defined through a Morse function $f$ which has
just one critical point per critical value can be decomposed into
elementary cobordisms, stacked together as $M_1$ and $M_2$ in figure
\ref{stack.fig}. By repeatedly applying lemma \ref{stacking.lemma} we obtain

\begin{corollary}\label{stacking.corollary} Let $(M,g)$ be a
 Morse geometry associated with a cobordism $\cM$ and a Morse
function $f:\cM\rar \mathR$ which has one critical point per
critical value. Then $(M,g)$ is causally continuous if and only if
each of the elementary Morse geometries into which it decomposes is
causally continuous.
\end{corollary}

\subsection{Inserting Morse neighbourhoods}\label{insert.section}

We start by showing that the converse of claim~\ref{IUcapcd.claim}
holds in an elementary Morse geometry $(E,g)$, when the open set in
question is a neighbourhood of the critical point.

\begin{lemma}\label{insert.lemma}
Let ${\cal E}$ be an elementary cobordism, $(E,g)$ be an associated
Morse geometry, and $Q\subset E$ be a neighbourhood of the Morse point such
that $I^\pm(x,Q) = I^\pm(x)\cap Q$ $\forall x \in Q$. If 
$(Q,g)$ is causally continuous, then $(E,g)$ too is causally
continuous.
\end{lemma}

Guided by the proof of the stacking lemma, we seek a neighbourhood
$Q'$ of $p$, an ``elsewhere region'' $S$ and a 
compact hypersurface $\Sigma$ which
satisfy: (i) $(Q',g)$ and $(S,g)$ are causally continuous,
$I^\pm(x,Q')=I^\pm(x)\cap Q'\;\forall\, x\in Q'$ and
$I^\pm(x,S)=I^\pm(x)\cap S\;\forall\, x\in S$; (ii) there is an
elementary cobordism $E' = f^{-1}([a,b])\subset E$ with $E'= Q'\cup
S$; (iii) the hypersurface $\Sigma$ is contained in $Q'\cap S$ and
separates $Q'-S\cap Q'$ and $S-Q'\cap S$. Then we can proceed as in
the proof of lemma~\ref{stacking.lemma}, with $Q'$ and $S$ playing the
role of $M_1$ and $M_2$ there.

\bproof Let $(Q,g)$ be causally continuous. Let $D_A\subset D_B\subset D_C$
be open $n$-discs centered at $p$ such that $\ovl{D_C}\subset Q$, with
$\nmion$-sphere boundaries $S_A$, $S_B$ and $S_C$, respectively.  We know
from the proof of claim~\ref{curvesconfined.claim} that there is an
$\alpha$ with $0<\alpha< min(c, 1-c)$ such that along any future-directed
causal curve $\gamma$ with an arc between $S_A$ and $S_B$, the Morse
function must increase by at least $\alpha$. Let
$\Pi=f^{-1}\left([0,c)\right)$ and $\Phi=f^{-1}\left((c,1]\right)$ be the
subcritical and supercritical regions in $E$, respectively. Thus if $x\in
D_A\cap \Pi$ and $y\in I^-(x)\cap (E-D_B)$, then $f(y)<c-\alpha$ and
similarly if $x\in D_A\cap \Phi$ and $y\in I^+(x)\cap(E-D_B)$ then $f(y)>
c+\alpha$. Now consider  $a=c-\alpha$ and $b=c+\alpha$, so that  $E'=
f^{-1}\left([a,b]\right)$ is  the associated thinner elementary Morse
geometry contained in $E$.

Define $Q'\equiv Q\cap E'$ and $S\equiv D(V_c,E')$, the domain of
dependence of the critical surface in $(E',g)$. 
Moreover, let  $\Sigma\equiv S_C\cap E'$. We now demonstrate that $Q'$, $S$ and
$\Sigma$ satisfy the conditions (i), (ii) and (iii) mentioned above, so
that the proof reduces to the proof of the stacking lemma:

(i) By construction, the set $Q'\subset Q$ satisfies $I^\pm(x,Q')=
I^\pm(x)\cap Q'\;\forall\, x\in Q'$. It follows from
claim~\ref{IUcapcd.claim} that $(Q',g)$ is causally continuous. The
set $S$ is convex and causally continuous, since $V_c$ is a closed
achronal subset of $E$ (see section~\ref{cs.section}).

(ii) To check that $E'= Q'\cup S$ we just need to verify that any
point $x\in E'-D_B\cap E'$ is in $S$, since $D_B\cap E'\subset Q'$. We
assume that $x\in \Phi$, the case with $x\in \Pi$ follows from dual
arguments. So suppose there is a past-inextendible causal curve
through $x$ which does not intersect $V_c$. Then, since
$J^-(x)\cap D_A=\emptyset$, it must be confined in the region
$f^{-1}([c,f(x)])- \left(D_A\cap f^{-1}([c,f(x)])\right)$, which is
homeomorphic to $V_1\times [0,1]$, where $V_1$ is the 
future boundary of $E$, and therefore compact. This is
impossible, since $(E,g)$ is strongly causal.

(iii) Finally, it is clear from the construction that $\Sigma=S_C\cap
E'$ is homeomorphic to $S^{\lamion}\times S^{\nmila-1}\times [0,1]$,
and therefore compact. Moreover $\Sigma$ separates $Q'-S\cap Q'$
from $S-Q'\cap S$.
 
We now demonstrate that $(E',g)$ is causally continuous. 
Causal continuity of $(E,g)$  will then follow, since $E$ can be expressed as
$M_1\cup E'\cup M_2$, with $M_1$ and $M_2$ being causally continuous product
cobordisms, and then  the stacking lemma applies.

We assume the contrary, i.e. that $(E',g)$ is causally discontinuous.
Thus, there exist  points $x$ and $y$ in $\inter{E'}$ satisfying 
$x\in\cpa{I^+(y)}-\ovl{I^-(y)}$.  Because $I^{\pm}(z,Q')=I^{\pm}(z)\cap
Q'\ \forall\, z\in Q$, claim~\ref{IUcap.claim} tells us that
$\scpa{Q'}{I^+(y,Q')}=\cpa{I^+(y)}\cap Q'$. Thus, if both $x$ and $y$ belong
to $Q'$ this would mean that $(Q',g)$ is causally discontinuous, which is a
contradiction. Similarly, the causal continuity of $S$ means that $x$ and
$y$ cannot both be in $S$.  Thus, $x\in Q'-S\cap S$ and $y\in S-Q'\cap S$
or vice-versa. Let us consider the first possibility, i.e., $x\in Q'-S\cap
Q'$. The other possibility  is covered by swapping $S$ and $Q'$ in the
argument that follows. 

Consider a sequence of points $y_k \rightarrow y$ with $y_k \in
I^+(y)$. There exists a sequence of timelike curves $\gamma_k$ from
$x$ to $y_k$, all of which intersect $\Sigma$, since by (iii) any
curve from a point $x$ in $Q'-S\cap Q'$ to a point $y$ in $S-Q'\cap S$
must cross $\Sigma$ at least once.  Let $z_k$ be the first point at
which each $\gamma_k$ intersects $\Sigma$.  A subsequence of
the $z_k$ converges to a point $z\in \Sigma$. From here one proceeds
exactly as in the proof of lemma \ref{stacking.lemma}, namely one uses
the causal continuity of $S$ and $Q'$ to deduce $x\in\ovl{I^-(y)}$, a
contradiction. \eproof

This result can be easily extended to the case where there are several
Morse points in the same critical surface. If each critical point
$p_j$ has a neighbourhood $Q_j$ such that $(Q_j,g)$ is causally
continuous and $I^\pm(x,Q_j)= I^\pm(x)\cap Q_j\;\forall\, x\in Q_j$,
then $(M,g)$ too will be causally continuous. Therefore, to establish
that every Morse geometry $(M,g)$ without index $1$ or $\nmion$
points is causally continuous it would suffice to find that around a
critical point of index $\lambda\neq 1,\nmion$ there is a
neighbourhood $Q$ such that $(Q,g)$ is causally continuous and $Q$ is
$I$-convex in $M$. 

The following claim guarantees that any
neighbourhood of a critical  point
contains neighbourhoods which are $I$-convex relative to
the whole Morse spacetime.

\begin{claim}\label{convexQ.claim} Let $(M,g)$ be a Morse geometry  
and $\cDe$ be a  round neighbourhood of a critical point $p\in \cM$, 
then there exists a punctured neighbourhood $Q\subset \cDe$ of $p$ which is 
$I$-convex with respect to $(M,g)$.
\end{claim}

\bproof Let $h$ be the Riemannian metric and $f$ the Morse function 
from which $g$ is constructed. From lemma
\ref{topeq.lemma} we know there is a homeomorphism 
$\Psi:\cDeprime\rar \cN$ between neighbourhoods $\cDeprime$, $\cN$ of
$p$ contained in $\cDe$, so that we know the topology of the ingoing
and outgoing basins of $\xi$. Let $c=f(p)$ be the critical value and
let the numbers $a$,$b$, with $a<c<b$, be such that the sets
$A_{\eta}\equiv V_a \cap \Deprimela$ and $B_{\eta}\equiv V_b\cap
\Deprimenmila$ (where $V_a \equiv f^{-1}(a)$, {\it etc.})
are respectively a $\lambda-1$ sphere and an
$\nmila-1$ sphere.
Their images $A_{\xi}= \Psi(A_{\eta})\subset \cNla$ and 
$B_{\xi}=\Psi(B_{\eta})\subset \cNnmila$ are also spheres.  
Define the set $Q= I^+(A_{\xi})\cap I^-(B_{\xi})$ in $M$. Then
$Q$ is $I$-convex in $M$. 

That $Q$ is a neighbourhood of $p$ can be immediately seen from the
topological equivalence between the $\eta$ and the $\xi$ flows in
$\cN$. We can ensure that $Q$ is contained in $\De$, by choosing the
numbers $a$ and $b$ close enough to $c$.  To see this, let $U$ be a
punctured neighbourhood of $p$ with $\ovl{U}\subset \De$.  From
claim~\ref{curvesconfined.claim} we know that there is a number
$\alpha_U$ such that for any $x\in U$ and $z\notin \De$ connected
through a timelike curve we must have $|f(x)-f(z)|>\alpha_U$. Let
$U'\subset U$ be small enough so that for any $x\in U'$ we have
$|f(x)-c|<\alpha_U$. We can assume that $N=\cN\mip\subset U'$. Define
then $a=c-\alpha$ and $b=c+\alpha$ for some small $\alpha< \alpha_U$
so that $A_{\eta}$, $B_{\eta}$ are spheres in $N$. Then $A_{\xi}$ and
$B_{\xi}$ are contained in $U'$ and therefore $I^+(A_{\xi})\cap
I^-(B_{\xi})$ is contained in $\De$. Otherwise, if there were points
$z\notin \De$, $x\in A_{\xi}$ and $y\in B_{\xi}$ with $z\in I^+(x)\cap
I^-(y)$, we would get the contradiction $c< f(x)+\alpha < f(z) < f(y)-
\alpha <c$. \eproof

We have established this result here because it may be needed in the
future, to complete the proof of the Borde-Sorkin
conjecture. We do not need it for our present
purposes, since directly from lemma~\ref{insert.lemma} we obtain:

\begin{corollary}\label{cccaninsertion.corollary} Let $\cE$ be an 
elementary cobordism and $f:\cE\rar [0,1]$ a Morse function with a
single Morse point $p$ of index $\la\neq 1, \nmion$. Let $g$ be a 
Morse metric constructed from $f$ and a Riemannian metric, $h$, 
which is Cartesian in a neighbourhood of the Morse point.  Then
$(E,g)$ is causally continuous.
\end{corollary}

Indeed, we know from previous work \cite{borde99} that the
 neighbourhood in which $h$ is Cartesian contains a
causally continuous neighbourhood Morse geometry $(\Qd,g)$
where $\Qd$ is $I$-convex. 
In fact, $\Qd$ has precisely the form $Q\cap
E'$ with $Q=I^+(\Aeta)\cap I^-(\Beta)$, as in
claim~\ref{convexQ.claim}, and $E'$ an open elementary cobordism in
$E$. 


\subsection{Causal continuity when $\lambda\neq 1, \nmion$}
\label{cclaneq1nmion.section}

Finally, we show that if the cobordism $\cM$ has a Morse function $f_0$
without index $1$ or $\nmion$ critical points, then there exist causally
continuous Morse spacetimes associated with $\cM$. 
We do this by constructing a Riemannian metric $h_0$ such that near the
critical points $\{p_k\}$, $h_0$ is Cartesian flat in precisely the same
coordinates in which $f_0$ takes the form (\ref{morselemma.eq}). 
We start by covering $\cM$ with a finite atlas $\{U_{\alpha}\}$ such that
the chart $U_k$ with $p_k\in U_k$ contains a round neighbourhood $D_k$ of
$p_k$ where the Morse function $f_0$ takes its canonical form and such that 
$D_k$
intersects none of the other charts. We construct 
the Riemannian metric $h_0$ using
an associated partition of unity $\{\theta_\alpha\}$ to patch together
local metrics:
\be
h_{0\mu\nu}(x) = \sum_\alpha \theta_\alpha(x) h^\alpha _{\mu\nu}(x)
\ee
When $\alpha = k$ corresponds to a chart containing critical 
point $p_k$, $h^k_{\mu\nu} = \delta_{\mu\nu}$. The other 
$h^\alpha_{\mu\nu}$ are arbitrary. With such an $h_0$, we 
can apply corollary \ref{cccaninsertion.corollary} and 
corollary~\ref{stacking.corollary}  and conclude
that $(M,g_0)$ is causally continuous. 

\section{Conclusions}\label{conclusions.section}

Our results suffice to establish the following proposition which, if weaker
than the Borde-Sorkin conjecture, suggests the same selection rule for
cobordisms in the gravitational Sum-Over-Histories.

\begin{proposition}\label{selection.proposition} Given a compact
cobordism $\cM$ and a Morse function $f:\cM\rar\mathR$ then: (i) if
$f$ has no critical points with Morse index $1$ or $\nmion$, there
exist Morse spacetimes $(M,g)$ associated with $\cM$ which are causally
continuous; (ii) if $f$ has critical points of index $1$ or $\nmion$,
every Morse geometry $(M,g)$ defined through $f$ is causally
discontinuous.
\end{proposition}

Another useful  way of stating this result is,  
\begin{corollary} 
If $\cM$ is a compact manifold admitting only Morse functions containing
critical points of index 1 or n-1 then it can support only causally
discontinuous Morse spacetimes. If $\cM$ admits a Morse
function which possesses no critical points of index 1 or n-1, then it can
support causally continuous Morse spacetimes.
\label{selection.corollary}
\end{corollary}

In \cite{surya97,dowker97}, we examined certain topologies corresponding to
important physical processes, namely, the pair production of black holes,
Kaluza Klein monopoles and geons.  We found that while the black hole and
Kaluza Klein cases admit Morse functions which have no index $1$ or $n-1$
points, this is not true for the pair production of irreducible geons.
Corollary \ref{selection.corollary} then tells us that the black hole and
monopole pair-production topologies in fact admit causally continuous Morse
histories, while the geon pair-production topologies do not. In
\cite{surya97} we moreover stated that there always exists a topological
transition between any two $3$ manifolds which admits a Morse function with
no index $1, n-1$ point, which means therefore that such a transition
always admits a causally continuous Morse history.

It of course remains to be seen if the Sorkin conjecture which  relates causal
continuity of a distinguishing spacetime to the non-singular propagation of
quantum fields on that background, can be verified in higher
dimensions. We leave this for future investigations. 

\section{Acknowledgments}

S.S. is grateful to Tomas Gomez for valuable help with Morse theory,
especially a key point in the proof of Lemma \ref{indexone.lemma}. We would
also like to thank G.Moore for help with dynamical flows. S.S. 
was supported by a postdoctoral fellowship from TIFR, Mumbai. F.D. 
is supported in part by an EPSRC Advanced Fellowship. R.G. is grateful
for support from the Beit Trustees.

\bibliographystyle{unsrt}
\bibliography{gmst}

\begin{thebibliography}{10}

\bibitem{gibbons92}
G.W. Gibbons and S.W. Hawking.
\newblock {\em Commun.Math.Phys.}, 148:345--352, 1992.

\bibitem{surya97}
H.F.Dowker and S.Surya.
\newblock {\em Phys. Rev. D}, D(58):124019, 1998.

\bibitem{dowker97}
H.F. Dowker and R.S. Garcia.
\newblock {\em Class. and Quant.Grav.}, 15:1859, 1998.
\newblock gr-qc/9711042.

\bibitem{borde99}
A.~Borde, H.F. Dowker, R.S. Garcia, R.D. Sorkin, and S.~Surya.
\newblock {\em Class. and Quant.Grav.}, 1999.
\newblock to appear.

\bibitem{hawking78a}
S.W. Hawking.
\newblock {\em Phys.Rev.D}, 18:1747--1753, 1978.

\bibitem{hawking78b}
S.W. Hawking.
\newblock {\em Nucl.Phys.B}, 144:349--362, 1978.

\bibitem{horowitz91}
G.T. Horowitz.
\newblock {\em Class. and Quant.Grav.}, {8}:587, 1991.

\bibitem{vilenkin94}
A.~Vilenkin.
\newblock {\em Phys.Rev. D}, 50:2581--2594, 1994.
\newblock gr-qc/9403010.

\bibitem{carlip98}
S.Carlip.
\newblock {\em Class. and Quant.Grav.}, {15}:2629, 1998.

\bibitem{geroch67}
R.P.Geroch.
\newblock {\em J. Math. Phys.}, {8}(4):782, 1967.

\bibitem{milnor65}
J.Milnor.
\newblock {\em Lectures on the h-cobordism theorem}.
\newblock Princeton University Press, Princeton, New Jersey, 1965.

\bibitem{ellis73}
S.W.Hawking and G.G.R.Ellis.
\newblock {\em The Large Scale Structure of Space-Time}.
\newblock Cambridge University Press, Cambridge, 1973.

\bibitem{sachs73}
S.W.Hawking and R.K.Sachs.
\newblock {\em Comm. Math. Phys.}, 35:287, 1973.

\bibitem{penrose72}
R.Penrose.
\newblock {\em Techniques of differential topology in general relativity}.
\newblock Society for Industrial and Applied Mathematics, Pennsylvania, 1973.

\bibitem{beem81}
J.K. Beem and P.E. Ehrlich.
\newblock {\em Global Lorentzian Geometry}.
\newblock Marcel Dekker, New York, 1981.

\bibitem{arnold92}
V.I.Arnol'd.
\newblock {\em Ordinary Differential Equations}.
\newblock Springer-Verlag, Berlin, 1992.

\bibitem{perko91}
L.Perko.
\newblock {\em Differential equations and dynamical systems}.
\newblock Springer-Verlag, New York, 1991.

\end{thebibliography}
\end{document}